\documentclass[conference]{IEEEtran}

\usepackage{amsmath}
\usepackage{amsfonts}
\usepackage{amssymb}
\usepackage{amsbsy}
\usepackage{graphicx}
\usepackage{times}
\usepackage{default}
\usepackage{color}
\usepackage[noadjust]{cite}
\usepackage{enumerate}
\usepackage{tikz}
\usepackage{subcaption}
\usepackage{setspace}
\usepackage{bibspacing}

\renewcommand{\tilde}{\widetilde}
\renewenvironment{pmatrix}{\lefto (\begin{matrix}}{\end{matrix}\right )}

\def\ba#1\ea{\begin{align*}#1\end{align*}}	
\def\ban#1\ean{\begin{align}#1\end{align}}	
\def\bac#1\eac{\vspace{\abovedisplayskip}{\par\centering$\begin{aligned}#1\end{aligned}$\par}\addvspace{\belowdisplayskip}}	

\newcommand{\lefto}{\mathopen{}\left}
\newcommand\mydots{\makebox[1em][c]{.\hfil.\hfil.}}

\newtheorem{theorem}{Theorem}

\newtheorem{lemma}{Lemma}

\newtheorem{proposition}{Proposition}

\newtheorem{remark}{Remark}

\IEEEoverridecommandlockouts 

\makeatletter

\newcommand{\vast}{\bBigg@{3}}
\newcommand{\Vast}{\bBigg@{4}}
\makeatother

\newcommand\blfootnote[1]{%
  \begingroup
  \renewcommand\thefootnote{}\footnote{#1}%
  \addtocounter{footnote}{-1}%
  \endgroup
}

\title{Necessary Conditions for \\ K/2 Degrees of Freedom }

\author{\IEEEauthorblockN{Recep G\"{u}l\IEEEauthorrefmark{1}, Helmut B\"{o}lcskei\IEEEauthorrefmark{1}, and Shlomo Shamai (Shitz)\IEEEauthorrefmark{2}}
\IEEEauthorblockA{\IEEEauthorrefmark{1}Dept. IT \& EE, ETH Z\"urich, Z\"urich, Switzerland \\
\IEEEauthorrefmark{2}Dept. EE, Technion-Israel Institute of Technology, Haifa, Israel
}}

\begin{document}
\maketitle

\begin{abstract}
Stotz et al., 2016, reported a sufficient (injectivity) condition for each user in a $K$-user single-antenna constant interference channel to achieve $1/2$ degree of freedom. The present paper proves that this condition is necessary as well and hence provides an equivalence characterization of interference channel matrices allowing full degrees of freedom. 
\end{abstract}
\vspace{-0.1 pt}
\section{INTRODUCTION}
Cadambe and Jafar \cite{Cadambe08, Jafar11} proposed a signaling scheme---known as interference alignment---that exploits time-frequency selectivity to achieve $K/2$ degrees of freedom (DoF) in $K$-user single-antenna interference channels (ICs). In \cite{Etkin09} and \cite{Motahari14} it was shown that $K/2$ DoF can also be achieved in ICs with constant channel matrix, i.e, in the absence of selectivity. Wu et al. \cite{Wu15} developed a general formula for the number of DoF in single-antenna ICs, extended to vector ICs in \cite{Stotz162}. This formula can, however, be difficult to evaluate as it is expressed in terms of R\'enyi information dimension \cite{Renyi59}. Building on the work by Wu et al. \cite{Wu15} and a recent breakthrough result in fractal geometry by Hochman \cite{Hochman14}, Stotz and B\"olcskei \cite{Stotz16} derived a DoF-formula for single-antenna ICs, which is exclusively in terms of Shannon entropy; this formula was then used to develop an explicit sufficient condition for  achieving $K/2$ DoF.  Stotz et al. \cite{Stotz163} later identified an even more general sufficient (injectivity) condition for each user to achieve $1/2$ DoF and hence $K/2$ DoF in total. \par 
 The main contribution of the present paper is to establish that the sufficient condition in \cite{Stotz163} is also necessary for each user to achieve $1/2$ DoF in fully connected ICs\footnote{Note that the sufficient condition in \cite{Stotz163} applies to all ICs, whereas here we restrict ourselves to fully connected ICs.}. The tools used in the proof of this result are the DoF-formula developed in \cite{Stotz16} and the entropy version of the Pl\"unnecke-Ruzsa inequality \cite{Tao10}.  
 \blfootnote{The work of S. Shamai (Shitz) was supported by the
European Union's Horizon 2020 Research And Innovation Programme,
grant agreement no. 694630.}
\section{SYSTEM MODEL}
We consider a single-antenna $K$-user (additive) IC with fully connected channel matrix $\mathbf{H}=(h_{ij})_{1\leqslant i,j\leqslant K}\in \mathbb{R}^{K\times K}$, i.e, $h_{ij} \neq 0$ for all $i,j$, and input-output relation 
\begin{equation*}
Y_{i}=\sqrt{\mathsf{snr}}\sum_{j=1}^{K}h_{ij}X_{j}+Z_{i},\,\, i=1,\!\mydots, K,
\end{equation*}
where $X_i \in \mathbb{R}$ is the input at the $i$-th transmitter, $Y_i \in \mathbb{R}$ is the output at the $i$-th receiver, and $Z_i \in \mathbb{R}$ is noise of absolutely continuous distribution satisfying $h(Z_i) > - \infty$ and $H(\lfloor Z_i \rfloor ) < \infty$. The input signals are independent across transmitters, and noise is i.i.d. across users and channel uses. \par
The channel matrix $\mathbf{H}$ is known perfectly at all transmitters and receivers.  We impose the average power constraint
 \begin{equation*} 
\frac{1}{n}\sum_{k=1}^{n}(x_{i}^{(k)})^{2}\leqslant 1
\end{equation*}
 on codewords $(x_{i}^{(1)} \mydots\text{ } x_{i}^{(n)})$ of block-length $n$ transmitted by user $i=1, \!\mydots, K$. 
 The number of DoF of the channel is 
 \begin{equation*}
\mathsf{DoF}(\mathbf{H}):=\mathop{\lim\sup}_{\mathsf{snr}\rightarrow\infty}\frac{\overline{C}(\mathbf{H};\mathsf{snr})}{\frac{1}{2}\log \mathsf{snr}},
\end{equation*}
where $\overline{C}(\mathbf{H};\mathsf{snr})$ stands for the sum-capacity of the IC. \par 
We note that \cite{Stotz163} studies conditions for each user to achieve $1/2$ DoF as opposed to $K/2$ DoF for all users in total. This is done to avoid trivial exceptions induced by particular network topologies reflected by zeros in the IC matrix. Here, we also study conditions for each user to achieve $1/2$ DoF. We remark, however, that topology-induced exceptions cannot occur for fully connected ICs.
 \section{MAIN RESULT}
We start by reviewing the sufficient (injectivity) condition for each user to achieve $1/2$ DoF identified in \cite{Stotz163}. Let $\mathbf{\check{h}}\in \mathbb{R}^{K(K-1)}$ denote the vector containing the off-diagonal elements of $\mathbf{H}$. There are 
 \begin{equation*} 
\varphi(d):=\begin{pmatrix}
K(K-1)+d\\
d
\end{pmatrix}
\end{equation*}
monomials\footnote{A monomial in $k$ variables $x_i$ is an expression of the form $x_1^{n_1}x_2^{n_2} \mydots \ x_k^{n_k}$, where $n_1,\! \mydots, n_k \in \mathbb{N}$, and the degree of the monomial is $n_1+\, \mydots+n_k$.} in $K(K-1)$ variables of degree not larger than $d$, which we enumerate as $f_{1}, \! \mydots, f_{\varphi(d)}$.
We will further need the following sets
\begin{equation*}
\mathcal{W}_{N,d}:=\Bigg\{\sum_{i=1}^{\varphi(d)}a_{i}f_{i}(\check{\mathbf{h}}):a_{1}, \! \mydots, a_{\varphi(d)}\in\{0,\! \mydots, N-1\}\Bigg\}
\end{equation*}
\begin{equation} \label{4} 
\mathcal{W}:=\bigcup_{d\geqslant 0}\bigcup_{N\geqslant 1}\mathcal{W}_{N,d}.
\end{equation}
It was shown in \cite[Th. 1]{Stotz163} that for each user to achieve $1/2$ DoF, it is sufficient that the following condition be satisfied for either $\mathbf{H}$ directly or at least one scaled version of $\mathbf{H}$, where scaling, as defined in \cite{Stotz163}, refers to a finite number of scalings of individual rows and columns by nonzero constants:  \par
For each $i=1, \! \mydots, K$, the map
\begin{equation*} \setlength{\abovedisplayskip}{8pt} \setlength{\belowdisplayskip}{7pt}
\qquad \qquad \qquad \qquad
\begin{matrix}
\mathcal{W}\times \mathcal{W}\rightarrow \mathcal{W}+h_{ii}\mathcal{W}\\[4pt]
(w_{1}, w_{2})\mapsto w_{1}+h_{ii}w_{2}
\end{matrix}\qquad\qquad \qquad 
(\ast \ast)
\end{equation*} 
is injective. \par
\vspace{0.6mm}
The main result of the present paper is as follows. 
\vspace{2.3mm} 
\begin{theorem}\label{thm1} 
Let $\mathcal{W}$ be as in (\ref{4}). Then, for almost all fully connected IC matrices $\mathbf{H}$, the following condition is necessary for each user to achieve $1/2$ DoF:  Either the channel matrix $\mathbf{H}$ itself or at least one scaled version thereof satisfies condition ($\ast \ast$). \par
\vspace{2pt}
\textit{Proof:}  See Section \ref{proofthm1}. 
\vspace{2pt}
\end{theorem} \par
With the matching sufficient condition in \cite[Th. 1]{Stotz163} this yields an equivalence characterization of full-DoF-achieving IC matrices.\vspace{2.2pt}
\begin{remark} \label{remarkChannel}
 Note that thanks to $\mathcal{W}$ containing integer linear combinations of monomials in the off-diagonal elements of $\mathbf{H}$, the condition ($\ast \ast$) is exclusively in terms of channel coefficients. 
\end{remark} \vspace{2.5pt}
\begin{remark} \label{remark1} 
Almost all channel matrices $\mathbf{H}$ are fully connected. Since an almost all subset of an almost all set is itself almost all, the set of channel matrices $\mathbf{H}$ not covered by Theorem \ref{thm1} has measure zero. 
\end{remark} 
\section{BALANCING RESULTS}
We start with a simple modification of \cite[Th. 3]{Stotz16}, with the proof omitted due to space constraints. \vspace{3pt}
\begin{proposition} \label{prop}
For almost all fully connected IC matrices $\mathbf{H}$ the following holds: If each user achieves 1/2 DoF, then for $\varepsilon \in (0,1/2)$, there exist independent discrete random variables $V_1, \! \mydots, V_K$ of finite entropy such that
\begin{align} \label{prop1}  \frac{1}{2}-\varepsilon \leq  \frac { \Big [{ H \Big ({\sum _{j=1}^{K} h_{ij} V_{j} }\Big )- H \Big ({ \sum _{j\neq i}^{K} h_{ij} V_{j} }\Big )}\Big ]}{ \max _{i=1, \mydots ,K} H   \Big ({\sum _{j=1}^{K} h_{ij} V_{j} }\Big )}, \end{align}
for $i=1, \! \mydots, K$, with the denominator of (\ref{prop1}) nonzero. \vspace{2mm}
\end{proposition} \par
The following lemma, based on Proposition \ref{prop}, states ``balancing'' results on the entropies of signal and interference contributions and will turn out instrumental in the proof of our main statement. \vspace{3pt}
\begin{lemma} \label{lem1}
For $\varepsilon \in (0,1/2)$, let $V_1, \! \mydots,V_K$ be the corresponding independent discrete random variables satisfying (\ref{prop1}) for the fully connected IC matrix $\mathbf{H}$. Then,
\begin{flalign}  \label{8}
& \frac{H \Big ({\sum _{j \neq i}^{K} h_{ij} V_{j} }\Big )}{  H  (V_i)} =1 + \mathcal{O}(\varepsilon) 
 \text{,  \space for{ } } i=1, \! \mydots, K,  \\[3.5mm] 
\label{9}
 & \frac{H   (V_i  )}{  H  (V_j)} =1 + \mathcal{O}(\varepsilon) \text{, \space for   }  i, j \in \{1,\! \mydots,K \}, i \neq j,\\[3.5mm]
 \label{22}
 & \frac{H \Big ({\sum _{j=1}^{K} h_{ij} V_{j} }\Big )}{H(V_i)}=2 + \mathcal{O}(\varepsilon)  \text{, \space for } i=1, \! \mydots, K. 
\end{flalign} 
\end{lemma} 
\begin{IEEEproof} 
Starting from (\ref{prop1}), we have  \vspace{2pt}
\begin{align}  \frac{1}{2}-\varepsilon   \leq \frac { \Big [{ H \Big ({\sum _{j=1}^{K} h_{ij} V_{j} }\Big )- H  \Big ({ \sum _{j\neq i}^{K} h_{ij} V_{j} }\Big )}\Big ]}{  H  \Big ({\sum _{j=1}^{K} h_{ij} V_{j} }\Big )}, \end{align}
 for $i=1, \! \mydots, K$.
Rearranging terms, we get
\begin{equation}  \label{12}
2 H \Bigg({\sum _{j \neq i}^{K} h_{ij} V_{j} }\Bigg) \leq (1+2 \varepsilon) \, H \Bigg ({\sum _{j=1}^{K} h_{ij} V_{j} }\Bigg ).
\end{equation}
Invoking the following inequality for independent discrete random variables $X$, $Y$ \cite[Ex. 2.14]{Cover06}
\begin{equation} \label{wki} H  (X+Y) \leq H  (X ) + H  ( Y), \end{equation}
 on the right-hand side (RHS) of (\ref{12}) yields
\begin{equation}  \label{13}
   (1-2\varepsilon) { H   \Bigg ({\sum _{j \neq i}^{K} h_{ij} V_{j} }\Bigg )} \leq (1+2 \varepsilon) \, H    ( V_i ).
\end{equation}
Next, we show that for fully connected ICs
\begin{equation}  \label{14}
H   \Bigg ({\sum _{j \neq i}^{K} h_{ij} V_{j} }\Bigg) \geq \frac{(1-2\varepsilon)}{(1+2\varepsilon)} \, H  (V_i).
\end{equation}  
To this end, w.l.o.g., we assume that $H(V_1) \geq H(V_2) \geq \mydots \geq H(V_K)$. Applying \cite[Ex. 2.14]{Cover06}
\begin{equation} \label{wki2}
H(\alpha X+\beta Y) \geq \max{\{H(X),H(Y)}\}
\end{equation} 
repeatedly for independent discrete random variables $X$ and $Y$, and arbitrary $\alpha$, $\beta$  $\in \mathbb{R} \setminus \{0\}$, with $X= \sum _{\substack{j \neq 1,i }}^{K} h_{ij} V_{j}$, $Y=V_1$, $\alpha=1$, $\beta= h_{i1}$, for $i=2, \! \mydots, K$, we obtain
\begin{equation}  \label{15}
(1-2\varepsilon) \, H(V_i) \leq (1-2\varepsilon) \ H(V_1) \leq (1+2\varepsilon) \, { H   \Bigg ({\sum _{j \neq i}^{K} h_{ij} V_{j} }\Bigg )}.
\end{equation}  
This proves (\ref{14}) for $i=2, \! \mydots, K$. The case $i=1$ is obtained as follows. First note that
 \begin{equation} \label{16}
(1-2\varepsilon) \, H(V_1) \leq  (1-2\varepsilon) \, { H   \Bigg ({\sum _{j \neq i}^{K} h_{ij} V_{j} }\Bigg )} \leq (1+2 \varepsilon) \, H   ( V_i  )\end{equation}  
 for all $i \neq 1$, where the first inequality is by (\ref{wki2}) and the second by (\ref{13}). Further, again by (\ref{wki2}), we have 
\begin{equation*}
(1+2 \varepsilon) \, H    ( V_i ) \leq (1+2\varepsilon) \, { H  \Bigg ({\sum _{j \neq 1}^{K} h_{ij} V_{j} }\Bigg )}, 
\end{equation*} 
for all $i \neq 1$, and inserting into (\ref{16}) establishes (\ref{14}) for $i=1$. We can now combine (\ref{13}) and (\ref{14}) to get
\begin{equation} \label{17}
 \frac{1-2\varepsilon}{1+2\varepsilon} \leq \frac{H   \Big ({\sum _{j \neq i}^{K} h_{ij} V_{j} }\Big )}{H(V_i)} \leq \frac{1+2\varepsilon}{1-2\varepsilon}
\end{equation}
for all $i$, which yields (\ref{8}). \par
To prove (\ref{9}), we again assume, w.l.o.g., that $H(V_1) \geq \mydots \geq H(V_K)$, and simply note that thanks to (\ref{16}) 
\begin{equation}  \label{18}
\frac{1-2\varepsilon}{1 +2 \varepsilon} \leq   \frac{H(V_K)}{H(V_1)} \leq \frac{H(V_i)}{H(V_j)} \leq \frac{H(V_1)}{H(V_K)} \leq \frac{1+2 \varepsilon}{1- 2 \varepsilon},
\end{equation}
for $i,j=1, \! \mydots, K$.  \par 
Finally, to establish (\ref{22}), we start by noting that
\begin{align} \label{23}
\frac{H  \Big ({\sum _{j=1}^{K} h_{ij} V_{j} }\Big )}{H \mathopen {} ( V_{i} )}  \geq \ &  \Big(\frac{1-2\varepsilon}{1+2\varepsilon} \Big)  \frac{H   \Big ({\sum _{j=1}^{K} h_{ij} V_{j} }\Big )}{H    \Big ({\sum _{j \neq i}^{K} h_{ij} V_{j} }\Big )}  \\ \notag \geq \ & \frac{2 (1-2\varepsilon)}{(1+2\varepsilon)^2},
\end{align}
owing to (\ref{17}) and (\ref{12}).
Using (\ref{wki}) and (\ref{17}), it follows that
\begin{equation} \label{24}  H   \Bigg ({\sum _{j=1}^{K} h_{ij} V_{j} }\Bigg ) \leq \Big( 1+ \frac{1+2\varepsilon}{1-2\varepsilon} \Big) H   ({ V_{i} }).
\end{equation}
Combining (\ref{24}) with (\ref{23}) then yields (\ref{22}).
\end{IEEEproof} \par \vspace{5pt}
We conclude this section by recording, for later use,  a simple variation of the entropy version of the Pl\"unnecke-Ruzsa inequality \cite{Tao10}. \vspace{5pt}
\begin{lemma}\text{[A simple variation of \cite[Th. 2.8.2]{Tao10}]}:\label{lemtao}
For $\varepsilon \in (0,1/2)$, assume that there exist independent discrete random variables $X,Y_1, \mydots, Y_m$, all of finite entropy, such that, 
\begin{equation} \label{tao1}
\frac{H(X+Y_i)}{H(X)} = 1+ \mathcal{O}(\varepsilon),
\end{equation}
for $i=1, \! \mydots ,m$. Then, for finite $m$, 
\begin{equation} \label{tao2}
 \frac{H(X+Y_1+\mydots + Y_m)}{H(X)} = 1+ \mathcal{O}(\varepsilon).
\end{equation}
\begin{IEEEproof}
In \cite[Th. 2.8.2]{Tao10}, set $\log{K_i}= H(X)\mathcal{O}(\varepsilon)$ for $i=1, \! \mydots, m$, and take the additive group $G$ as $\mathbb{R}$. This results in $H(X+Y_1+\mydots + Y_m) \leq H(X)+ m H(X) \mathcal{O}(\varepsilon) $. Divide this inequality by $H(X)$ and note that owing to (\ref{wki2}) the expression on the LHS of (\ref{tao2}) is greater than or equal to 1. The proof is concluded by noting that $\mathcal{O}(\varepsilon) m =\mathcal{O}(\varepsilon)$ for finite $m$. 
\end{IEEEproof}
\end{lemma}
\vspace{5pt}
\section{PROOF OF THEOREM \ref{thm1}} \label{proofthm1}
For simplicity of exposition and due to space constraints, we detail the proof for the 3-user case only. The proof for the $K$-user case follows by induction over the number of users with the $3$-user case constituting the base case.  \par 
Consider the 3-user fully connected IC matrix
\begin{equation*}  \mathbf{H}=\begin{pmatrix} h_{11}&h_{12}&h_{13}\\ h_{21}&h_{22}&h_{23}\\ h_{31}&h_{32}&h_{33} \end{pmatrix}\hspace{-2.6pt}. \end{equation*} 
The proof is effected by contradiction, with the contradiction established by induction on the degrees of polynomials in $\mathcal{W}$ as defined in (\ref{4}). Towards this contradiction, we assume that $\mathbf{H}$ is in the almost all set covered by Theorem \ref{thm1}, while at the same time each user achieves $1/2$ DoF and condition ($\ast \ast$) is violated for $\mathbf{H}$ and all scaled versions thereof. In particular, condition ($\ast \ast$) must also be violated for 
\begin{equation} \label{5} \mathbf{\tilde{H}}=\begin{pmatrix} g_1&1&1\\ 1&g_2&1\\ 1&h&g_3 \end{pmatrix} \hspace{-2pt},\text{\hspace{-2.7pt}$ \hspace{1.3pt}\ g_1, g_2, g_3, h \neq 0$,}\end{equation}
which can be obtained from $\mathbf{H}$ by scaling (cf. \cite[p. 259]{Wu15}). It follows from a simple modification of the arguments in \cite[Lem. 1]{Etkin09} that for fully connected ICs, scaling does not change the number of DoF achieved by each user. Hence, each user achieves $1/2$ DoF in $\mathbf{\tilde{H}}$ as well. The reduction to $\mathbf{\tilde{H}}$ is for simplicity of exposition as all elements of the corresponding set $\mathcal{W}$ in (\ref{4}) are polynomials in $h$.  \par 
We proceed by assuming that the map in ($\ast \ast$) is not injective for user $i=1$ so that condition ($\ast \ast$) is violated. This implies the existence of $w_1,w_2,\tilde{w}_1, \tilde{w}_2 \in \mathcal{W}$ such that $w_1 \neq \tilde{w}_1$, $w_2 \neq \tilde{w}_2 $, and 
\begin{equation*} 
w_1+ g_1 w_2 = \tilde{w}_1+ g_1 \tilde{w}_2. 
\end{equation*}
We can hence write 
\begin{equation} \label{43}
g_1=\frac{w_1-\tilde{w}_1}{\tilde{w}_2 -w_2 }=\frac{\sum_{p=0}^{d_1}{\hat{a}_p h^p }}{\sum_{p=0}^{d_2}{\hat{b}_p  h^p}},
\end{equation}
where $d_1$ and $d_2$ are finite and the coefficients $\hat{a}_p, \hat{b}_p$ are integers. Since $w_1 \neq \tilde{w}_1$ and $w_2 \neq \tilde{w}_2$, the polynomials on the RHS of (\ref{43}) are nonzero. \par
Since $\mathbf{\tilde{H}}$ is fully connected, we know from (\ref{22}) that, for $i=1$,
\begin{equation} \label{27}
 \frac{H   \Big(\frac{\sum_{p=0}^{d_1}{\hat{a}_p h^p}}{\sum_{p=0}^{d_2}{\hat{b}_p h^p}} V_1+ V_2 + V_3\Big) }{H(V_1)}=2+\mathcal{O}(\varepsilon).
\end{equation} 
We shall show that this leads to a contradiction, by proving that (\ref{prop1}) implies
\begin{equation} \label{28}
 \frac{H   \Big(\frac{\sum{_{p=0}^{d_1}a_p h^p}}{\sum_{p=0}^{d_2}{b_p h^p}} V_1+ V_2 + V_3\Big) }{H(V_1)}=1+ \mathcal{O}(\varepsilon)
\end{equation}
for every nonzero finite-degree polynomial in $h$ with integer coefficients $a_p$ and $b_p$, in particular $\hat{a}_p$ and $\hat{b}_p$. The implication (\ref{prop1}) $\implies$ (\ref{28}) is established by induction over $d= \max \{d_1,d_2\}$.  \vspace{4pt}\par
\textit{Base case ($d=0$)}: When $d=0$, we need to show that 
\begin{equation} \label{30}
 \frac{H   \Big( a_0 V_1 + b_0  (V_2+V_3) \Big)}{H(V_1)}=1+\mathcal{O}(\varepsilon).
\end{equation}
From (\ref{8}) with $i=1$ we have
\begin{equation} \label{25}
\frac{H  ( V_2 + V_3 )}{H(V_1)}=1+\mathcal{O}(\varepsilon).
\end{equation}
Likewise, (\ref{8}) with $i=2$, upon using (\ref{9}) with  $i=1$, $j=2$ yields
\begin{equation} \label{31}
\frac{H ( V_1 + V_3)}{H(V_1)}=1+\mathcal{O}(\varepsilon).
\end{equation}
As the polynomials in (\ref{28}) are nonzero, it suffices to prove the statement for $a_0,b_0 \in \mathbb{Z} \setminus \{0\}$. Using (\ref{9}) with $i=1$, $j=3$, we can replace the denominators of (\ref{25}) and (\ref{31}) by $H(V_3)$ and then apply Lemma \ref{lemtao} to get
\begin{equation} \label{44} 
\frac{H  (V_1+V_2+V_3)}{H(V_3)} =1+ \mathcal{O}(\varepsilon).
\end{equation}
Replacing $H(V_3)$ in (\ref{44}) by $H(V_1)$, which is possible thanks to (\ref{9}) with $i=1$, $j=3$, we obtain
\begin{equation} \label{45} 
\frac{H   ( V_1 + V_2+ V_3 )}{H(V_1)}=1+ \mathcal{O}(\varepsilon).
\end{equation}
Applying \cite[Th. 14]{Wu15} with $p=a_0, q=b_0, X=V_1$, and $Y=V_2+V_3$, and dividing the result thereof by $H(V_1)$ yields
 \begin{align}\label{47} 
\frac{H(a_0V_1+b_0 (V_2+V_3))}{H(V_1)}  -\frac{ H(V_1+V_2+V_3)}{H(V_1)} \leq & \\ \notag \tau_{a_0,b_0} \bigg (\frac{(2 H(V_1+V_2+V_3)-H(V_1)-H(V_2+V_3))}{H(V_1)} \bigg ),
\end{align}
where $\tau_{a_0,b_0}= 7 \lfloor \log |a_0| \rfloor + 7 \lfloor \log |b_0| \rfloor +2$.
By (\ref{25}) and (\ref{45}), the RHS of (\ref{47}) equals $\mathcal{O}(\varepsilon)$. We therefore have
\begin{equation} \label{48} 
1 \leq \frac{H   (a_0V_1+b_0 (V_2+V_3))}{H(V_1)}  \leq 1 +\mathcal{O}(\varepsilon),
\end{equation}
where the first inequality follows from (\ref{wki2}). In summary, we have
 \begin{equation} \label{49} 
 \frac{H  ( a_0 V_1 + b_0( V_2+V_3))}{H(V_1)}=1+ \mathcal{O}(\varepsilon),
 \end{equation}
 which establishes (\ref{30}) as desired.  \par \vspace{2pt}
 \textit{Induction step}. We assume that (\ref{28}) holds for $d = m-1$, with $m \geq 1$, and show that this implies (\ref{28}) for $d=m$. The following lemma contains the central technical result in the induction step. \vspace{2pt}
\begin{lemma} \label{lemCaseM}
 For $\varepsilon \in (0,1/2)$, let $V_1,V_2$, and $V_3$ be the corresponding independent discrete random variables satisfying (\ref{prop1}) for the IC matrix $\mathbf{\tilde{H}}$ in (\ref{5}). Assume further that $V_1,V_2$, and $V_3$ satisfy (\ref{28}) for $d=\max{\{d_1,d_2\}}= m-1$, where $m \geq 1$. Let $V_1^*$ be an independent copy of $V_1$, and $\tilde{V}_2, \hat{V_2}$ and $\tilde{V}_3, \hat{V_3}$ independent copies of $V_2$ and $V_3$, respectively. Then,
\begin{flalign} \label{34}
 &\frac{H   \Big( h(V_2+V_3)+a_mh^mV_1^*+b_mh^m(\tilde{V}_2+\tilde{V}_3) \Big)}{ H(V_1)} = 1 +\mathcal{O}(\varepsilon) \\
 &\notag \frac{H   \Big( h(V_2+V_3)+\sum_{i=0}^{m-1}{a_i h^i} V_1 + {\sum_{i=0}^{m-1}{b_i h^i}} (\hat{V}_2+\hat{V}_3) \Big) }{ H(V_1)} =\\  \label{35} & 1+\mathcal{O}(\varepsilon)  \\
\label{504}
 & \frac{ H   \Big( \alpha_1 V_1^*+ \alpha_2 V_1 + \beta_1 (\tilde{V}_2+\tilde{V}_3)+ \beta_2 (V_2+V_3) \Big) }{H(V_1)}  +\mathcal{O}(\varepsilon) \geq  \\ &  \notag \frac{ H   \Big( (\alpha_1+ \alpha_2) V_1 + (\beta_1+\beta_2) (V_2+V_3) \Big) }{H(V_1)} ,
\end{flalign}
for all $\alpha_1, \alpha_2, \beta_1, \beta_2 \in \mathbb{R}$. \vspace{6pt}
\begin{IEEEproof}
To establish (\ref{34}), we apply Lemma \ref{lemtao} with $X=a_m h^m V_1^*$, $Y_1=h(V_2+V_3)$, and $Y_2= b_m h^m (\tilde{V}_2+\tilde{V}_3)$. The corresponding conditions (\ref{tao1}) hold as dividing $X+Y_1$ and $X+Y_2$ by $h$ yields (\ref{28}), which is satisfied by the induction hypothesis. \par 
For (\ref{35}) we have to distinguish the cases $m>1$ and $m=1$. For $m>1$, one applies Lemma \ref{lemtao} with $X=\sum_{i=0}^{m-1}{a_i h^i} V_1$, $Y_1=h(V_2+V_3)$, and $Y_2=\sum_{i=0}^{m-1}{b_i h^i}(\hat{V}_2+\hat{V}_3) $. Again, the corresponding conditions (\ref{tao1}) of Lemma \ref{lemtao} are satisfied thanks to the induction hypothesis. For $m=1$, (\ref{35}) reduces to
\begin{equation} \label{54}
  \frac{H  (h(V_2+ V_3) +a_0V_1+b_0(\hat{V}_2+ \hat{V}_3))}{H(V_1)} =1+\mathcal{O}(\varepsilon).
\end{equation}
Using (\ref{wki2}) and (\ref{45}), we get
\begin{equation} \label{60}
\frac{H(V_2)}{ H(V_1)} \leq \frac{H(h\hat{V}_2+h\hat{V}_3)}{ H(V_1)} \leq  \frac{H(V_1+V_2+V_3)}{ H(V_1)}=1+\mathcal{O}(\varepsilon).
\end{equation}
Thanks to (\ref{9}), the leftmost term in (\ref{60}) equals $1+\mathcal{O}(\varepsilon)$. Hence, 
\begin{equation} \label{500}
  \frac{H(h\hat{V}_2+h\hat{V}_3)}{ H(V_1)}=1+\mathcal{O}(\varepsilon).
\end{equation}
Furthermore, applying (\ref{8}) with $i=3$, we have  
\begin{equation} \label{61}
 \frac{H(h\hat{V}_2+V_1)}{ H(V_3)}=1+\mathcal{O}(\varepsilon),
\end{equation}
which thanks to (\ref{9}) with $i=1, j=3$ yields
\begin{equation} \label{501}
 \frac{H(h\hat{V}_2+V_1)}{ H(V_1)}=1+\mathcal{O}(\varepsilon).
\end{equation}
Combining (\ref{500}) and (\ref{501}), and applying Lemma \ref{lemtao} with $X=h \hat{V_2}$, $Y_1=V_1$, and $Y_2=h\hat{V_3}$, we obtain
\begin{equation} \label{62}
\frac{H(V_1+h(\hat{V}_2+\hat{V}_3))} {H(V_1)}=1+\mathcal{O}(\varepsilon).
\end{equation}
Repeating the steps leading from (\ref{45}) to (\ref{49}) now results in
\begin{equation} \label{63}
\frac{H(a_0V_1+b_0\hspace{0.1mm} h(\hat{V}_2+\hat{V}_3))}{H(V_1)}=1+\mathcal{O}(\varepsilon).
\end{equation}
\begin{spacing}{1.1} Using (\ref{63}) and applying Lemma \ref{lemtao} with $X=a_0 V_1$, $Y_1=b_0\hspace{0.1mm} h (\hat{V_2}+\hat{V_3})$, and $Y_2=h(V_2+V_3)$, we get \end{spacing} \vspace{-9pt}
\begin{equation} \label{65}
 \frac{H(h(V_2+V_3)+ a_0V_1+b_0 \hspace{0.1mm} h (\hat{V}_2+\hat{V}_3))}{H(V_1)}=1 +\mathcal{O}(\varepsilon),
\end{equation}
which establishes (\ref{54}), and thereby completes the proof of (\ref{35}).  \par
 To prove (\ref{504}), we first apply Lemma \ref{lemtao} with $X=V_1$, $Y_1=V_2+V_3$ and $Y_2= \tilde{V}_2+ \tilde{V}_3$, resulting in
\begin{equation} \label{505}
\frac{H(V_1+V_2+V_3+\tilde{V}_2+\tilde{V}_3)}{H(V_1)}=1+\mathcal{O}(\varepsilon).
\end{equation}
The corresponding conditions (\ref{tao1}) of Lemma \ref{lemtao}  are satisfied thanks to  (\ref{45}), and upon noting that $\tilde{V}_2, \tilde{V}_3$ are independent copies of $V_2, V_3$. Thanks to (\ref{25}) and (\ref{wki2}), we have
\begin{align} \label{506}
1+\mathcal{O}(\varepsilon)=\frac{H(V_2+V_3)}{H(V_1)}\leq   \frac{H(V_2+V_3+\tilde{V}_2+\tilde{V}_3)}{H(V_1)} \\ \leq  \frac{H(V_1+V_2+V_3+\tilde{V}_2+\tilde{V}_3)}{H(V_1)},
\end{align}
which, when combined with (\ref{505}), yields
\begin{equation}  \label{507}
 \frac{H(V_2+V_3+\tilde{V}_2+\tilde{V}_3)}{H(V_1)}=1+\mathcal{O}(\varepsilon).
\end{equation}
The same line of reasoning, with Lemma \ref{lemtao} applied with $X=V_3$, $Y_1=V_1$, and $Y_2= V_1^*$, upon invoking (\ref{31}), delivers
\begin{equation} \label{508}
\frac{H(V_1+V_1^*)}{H(V_1)}= 1+\mathcal{O}(\varepsilon).
\end{equation}
We next use \cite[Lem. 18]{Wu15} with $Z=\beta_1 (\tilde{V}_2+\tilde{V}_3)+ \beta_2 (V_2+V_3)$, $X= V_1$, $X'= V_1^*$, $r=\alpha_1$, and $p=\alpha_1+\alpha_2$ to get\footnote{Note that \cite[Lem. 18]{Wu15} assumes $p$ and $r$ to be nonzero integers. A closer inspection of the proof reveals, however, that the result holds for all $p,r \in \mathbb{R}$.} 
\begin{flalign} \label{509}
\notag \frac{ H   \Big(\! (\alpha_1 \!+\! \alpha_2)V_1+\beta_1 (\tilde{V}_2 \!+ \! \tilde{V}_3)+ \beta_2 (V_2 \! + \! V_3) \! \Big) }{H(V_1)} \! - \! \frac{\Delta (V_1, V^*_1)}{H(V_1)} & \\  \leq \      \frac{ H   \Big( \alpha_1 V_1^*+ \alpha_2 V_1 + \beta_1 (\tilde{V}_2+\tilde{V}_3)+ \beta_2 (V_2+V_3) \Big) }{H(V_1)} & ,
\end{flalign}
where $\Delta(V,W)=H(V-W)-\frac{1}{2}H(V)-\frac{1}{2}H(W)$.  
Note that due to (\ref{508}), we have 
\begin{equation} \label{510}
\frac{\Delta (V_1,-V^{*}_1)}{H(V_1)}=\mathcal{O}(\varepsilon).
\end{equation}
Thanks to \cite[Th. 3.5]{Madiman14} this implies 
\begin{equation} \label{511}
\frac{\Delta (V_1,V^*_1)}{H(V_1)}=\mathcal{O}(\varepsilon).
\end{equation} \begin{spacing}{1.05}
We next apply \cite[Lem. 18]{Wu15} with $Z=(\alpha_1+\alpha_2)V_1$, $X=V_2+V_3$, $X'=\tilde{V}_2+\tilde{V}_3$, $p=\beta_1+\beta_2$, and $r=\beta_2$. Noting that thanks to (\ref{507}), (\ref{510}) continues to hold if we replace $\Delta (V_1, -V_1^*)$ by $\Delta (V_2+V_3, -(\tilde{V}_2+\tilde{V}_3) )$, we obtain \end{spacing} \vspace{-15pt}
\begin{align} \label{512} \notag
& \frac{ H   \Big( (\alpha_1+ \alpha_2) V_1 + (\beta_1+\beta_2) (V_2+V_3) \Big) }{H(V_1)} \leq \mathcal{O}(\varepsilon) \\ & + \frac{ H   \Big( (\alpha_1 + \alpha_2)V_1+\beta_1 ({V}_2+{V}_3)+ \beta_2 (\tilde{V}_2+\tilde{V}_3) \Big) }{H(V_1)}. 
\end{align}
Finally, combining (\ref{509}) and (\ref{512}), we get (\ref{504}).
\end{IEEEproof}
\end{lemma} \par \vspace{3pt}
We now use Lemma \ref{lemCaseM} to finalize the induction argument. First, let 
\begin{align*}
p(h):= &\, h \, (V_2+V_3)+  \sum_{i=0}^{m-1}{a_i h^i} \hat{V}_1 + {\sum_{i=0}^{m-1}{b_i h^i}} (\hat{V}_2+\hat{V}_3) \\ & + a_mh^m V_1^*+b_mh^m(\tilde{V}_2+\tilde{V}_3),
\end{align*}
where $\tilde{V}_2, \hat{V_2}$ and $\tilde{V}_3, \hat{V_3}$ are independent copies of $V_2$ and $V_3$, respectively.
\par 
\begin{spacing}{1.07} Applying Lemma \ref{lemtao} with $X=h(V_2+V_3)$, $Y_1=a_mh^mV_1^*+ b_mh^m(\tilde{V}_2+\tilde{V}_3)$, and $Y_2= \sum_{i=0}^{m-1}{a_i h^i} \hat{V}_1 + {\sum_{i=0}^{m-1}{b_i h^i}} (\hat{V}_2+\hat{V}_3)$, noting that the corresponding conditions (\ref{tao1}) are satisfied thanks to (\ref{34}) and (\ref{35}), and invoking (\ref{31}), we get \end{spacing}
\begin{equation} \label{503}
 \frac{H(p(h))}{ H(V_1)}= 1+\mathcal{O}(\varepsilon).
\end{equation}
Now define 
\begin{align*}
\tilde{p}(h): = &\sum_{i=0}^{m-1}{a_i h^i} \hat{V}_1 + {\sum_{i=0}^{m-1}{b_i h^i}} (\hat{V}_2+\hat{V}_3) \\ &+ a_mh^m V_1^*+ b_mh^m(\tilde{V}_2+\tilde{V}_3).
\end{align*}
Thanks to (\ref{wki2}) and (\ref{503}), we have
\begin{equation} \label{514}
  1+\mathcal{O}(\varepsilon) =\frac{H(p(h))}{ H(V_1)} \geq \frac{H(\tilde{p}(h))}{ H(V_1)} \geq 1.
\end{equation}
Applying (\ref{504}) with $\alpha_1=\sum_{i=0}^{m-1}{a_i h^i}$, $\alpha_2=a_mh^m$, $\beta_1= {\sum_{i=0}^{m-1}{b_i h^i}}$, and $\beta_2=b_mh^m$, yields
\begin{equation} \label{513}
  \frac{H (\tilde{p}(h))}{ H(V_1)} +\mathcal{O}(\varepsilon) \geq \frac{H   \Big( \sum_{i=0}^{m}{a_i h^i} V_1 + {\sum_{i=0}^{m}{b_i h^i}} (V_2+V_3) \Big)}{ H(V_1)}.
\end{equation}
Thanks to (\ref{wki2}), the RHS of (\ref{513}) cannot be smaller than one. Combining (\ref{514}) and (\ref{513}), we hence get 
\begin{equation}
\frac{H  \Big( \sum_{i=0}^{m}{a_i h^i} V_1 + {\sum_{i=0}^{m}{b_i h^i}} (V_2+V_3) \Big)}{ H(V_1)}=1+\mathcal{O}(\varepsilon).
\end{equation} 
This concludes the induction argument and thereby the proof.

\setlength{\bibitemsep}{.2\baselineskip plus .05\baselineskip minus .05\baselineskip}
\bibliographystyle{IEEEtran}
\bibliography{Zitat}
\end{document}